# Anomalous magnetization behavior in $Ce(Fe,Si)_2$

Arabinda Haldar[1], K. G. Suresh[1,*] and A. K. Nigam[2]
[1]Magnetic Materials Laboratory, Department of Physics
Indian Institute of Technology Bombay, Mumbai- 400076, India
[2]Tata Institute of Fundamental Research,
Homi Bhabha Road, Mumbai- 400005, India

*Abstract*

We report the effect of Si doping on the magnetization behavior of $CeFe_2$. It is found that Si stabilizes the dynamic antiferromagnetic state of $CeFe_2$. Multi-step magnetization behavior, unusual relaxation effect, thermal and magnetic history dependence, which are signatures of martensitic scenario, are found to be present in this system. We also show that one can induce the magnetization steps with the help of appropriate measurement protocol. Detailed magnetization relaxation studies have been carried out to understand the dynamics of magnetic phase transition.

----------------------------------------------------------------
[*]Corresponding author (email: suresh@phy.iitb.ac.in)

## I. Introduction

Multiple magnetization steps seen in a few intermetallics compounds and a large number of oxides has drawn a lot of attention recently. It is well known that structural heterogeneities arising from the magnetic heterogeneities presents a martensitic-like scenario in these systems and is responsible for the anomalous magnetic behavior. A lot of research activity has been made to explore the origin of the martenistic scenario, especially in the intermetallics. This feature is found in single crystalline samples, thin films and even in polycrystalline compounds. This observation is quiet unusual in polycrystalline samples and attracts a lots of attention. Various explanations had been put forward with detailed experimental results like relief of strain across the phase separated region [1-4], burst like growth of ferromagnetic phase (FM) with increasing field [5, 6] and quenched disorder [7].

$CeFe_2$ is unique among the series of R (rare earth)-$Fe_2$ compounds. It has low saturation moment ($M_s$) $2.3\mu_B$ / f.u. ($M_s$= $2.9\mu_B$ / f.u. of $LuFe_2$), anomalously low Curie temperature ($T_C$); 230K ($T_C$=610K & 545K for $LuFe_2$ and $YFe_2$ respectively) [8, 9]. It is to be mentioned here that this system shows two phase nature when selected elements (Ru, Re, Al, Co, Ir, Os and Ga) is substituted at the iron site [10, 11]. Most interestingly sharp jumps in magnetization have been observed in magnetization in Ru, Re [3] and Ga [4] doped compounds. Unusual steps in low temperature magnetization isotherms in certain doped $CeFe_2$ compounds [3, 4] proved that the system can be a promising material to study the underlying physics in phase separated systems.

In this report we present a detailed study on temperature, field, time dependence on Si doped $CeFe_2$ with various Si concentrations. Effect of zero field cooling (ZFC) and field cooling (FC) the compound during the measurement is demonstrated. We have also studied relaxation phenomenon in detail to understand the dynamics of the steps observed in M(H) isotherms. The presence of induction time to occur these steps has been discussed.

## II. Experimental Details

Polycrystalline compounds $Ce(Fe_{1-x}Si_x)_2$ with x=0.01, 0.025 & 0.05 were prepared by arc melting of the constituent elements: Ce(99.9%), Fe(99.999%) and Si(99.999%). The details of preparation have been reported elsewhere [4]. The structural analysis was performed by taking room temperature x-ray diffraction using Cu-K$\alpha$ radiation. The diffractograms were refined by the Rietveld analysis using *Fullprof* suite program. Magnetization measurements in the temperature range of 1.8- 300 K and in fields up to 90 kOe were carried out in Physical Property Measurement System (PPMS, Quantum Design Model 6500) which has a vibrating sample magnetometer (VSM) attachment. Magnetization measurements have been taken both in zero field cooled (ZFC) and field cooled (FC) modes.

## III. Results and Discussions

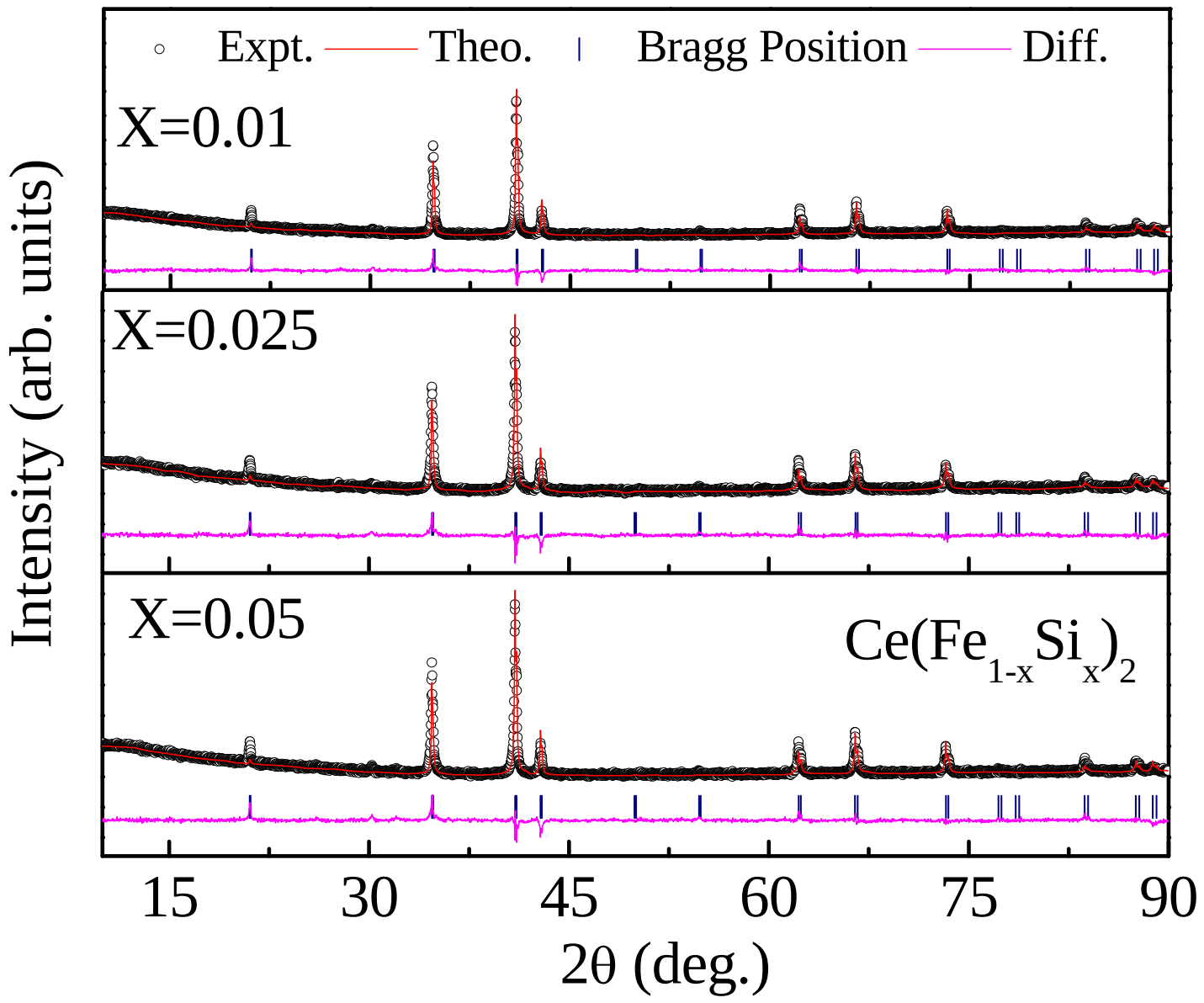

FIG. 1. Room temperature x-ray diffractograms for $Ce(Fe_{1-x}Si_x)_2$ compounds using Cu-K$\alpha$ radiation.

X-ray diffraction using Cu-K$\alpha$ radiation was taken on the samples with x=0.01, 0.025 & 0.05, compounds at room temperature. In fig. 1 open circles represents experimental points and the red line shows the fitted line obtained from Rietveld refinement of the diffractograms. As can be seen, all the compounds have formed in single phase. Rietveld refinement confirms the $MgCu_2$ type cubic structure with the space group of $\boldsymbol{Fd\bar{3}m}$ of all these compounds.

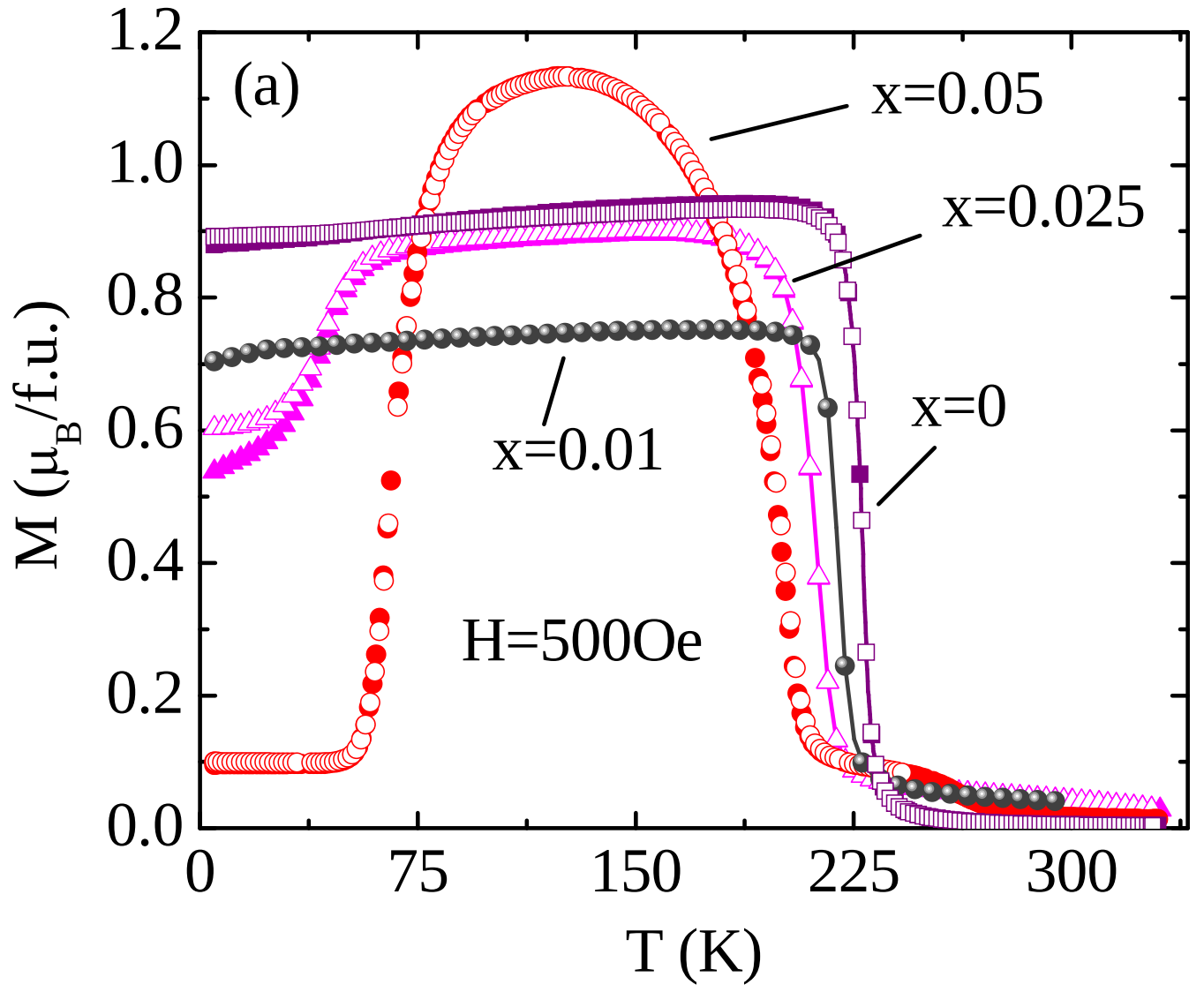

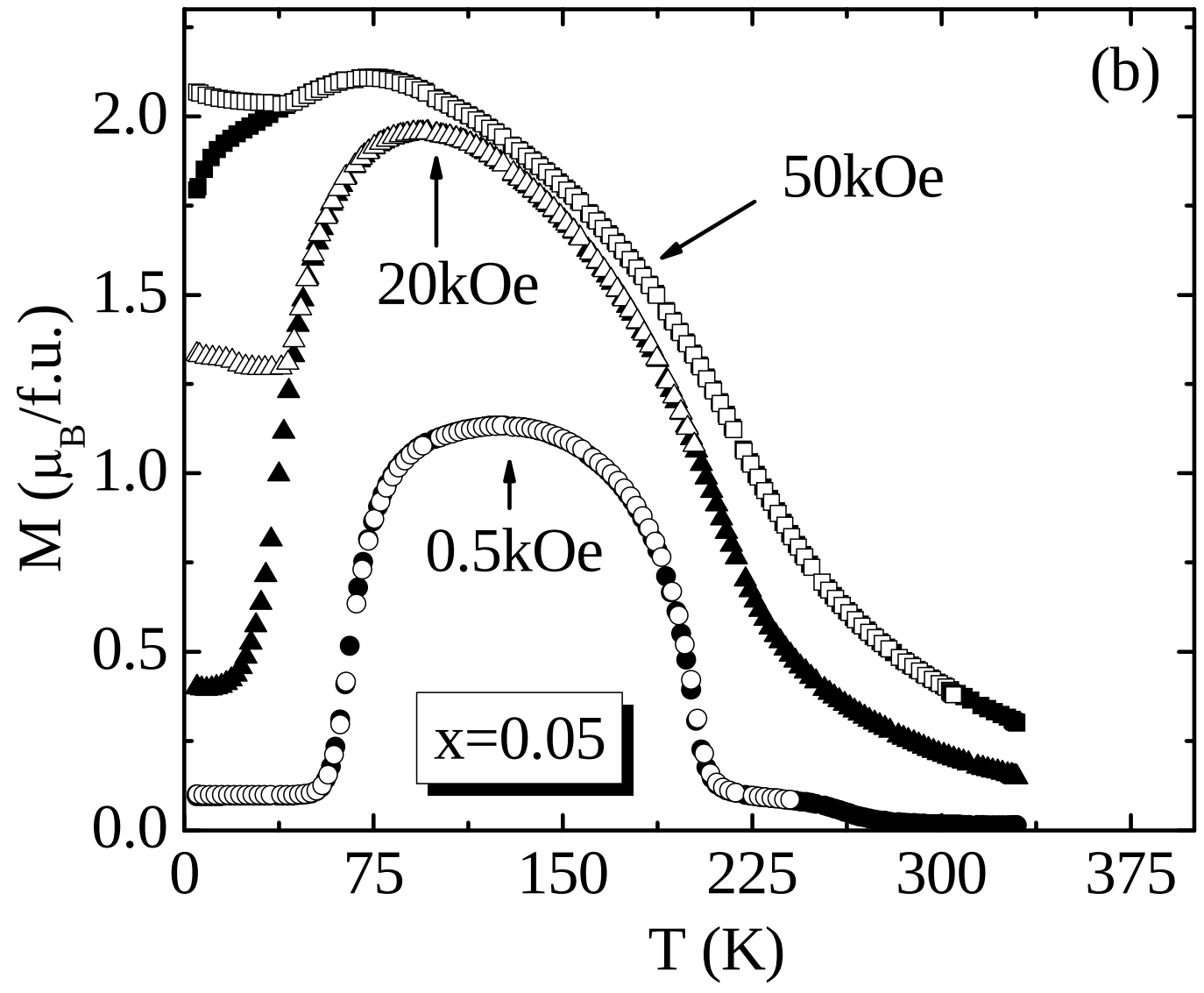

FIG. 2. (a) Temperature variation of magnetization of $Ce(Fe_{1-x}Si_x)_2$ compounds. (b) Temperature dependence of magnetization of $Ce(Fe_{1.95}Si_{0.05})_2$ compound at different applied magnetic fields. In both the cases data have been taken during warming the sample in both ZFC (closed points) and FC (open points) mode.

Fig. 2a shows the variation of magnetization with temperature of the all the compounds in a field of 500 Oe. M *vs*. T plot for $CeFe_2$ [4] has been shown for better comparison. With substitution of Si, the fluctuating antiferromagnetic (AFM) state present in the parent compound gets stabilized for x=0.025 & 0.05, which is similar to our recent result with Ga substitution [4]. For x=0.01 substitution Curie temperature decreases by 10 K compared to parent compound ($T_C$=230K) and with more substitution we have order-order (FM-AFM) transition at 41K (x=0.025) and 65K (x=0.05). The Curie temperature decreases monotonically with Si. As can be observed from the figure, AFM ground state is not fully stabilized in x=0.025 compounds whereas it is fully stabilized at x=0.10. From the ZFC and FCW data, there is no considerable difference observed between these two modes for x=0 and 0.1, while x=0.025 compound shows some difference at low temperatures. But x=0.05 compound shows a huge difference between ZFC and FC data at low temperatures. This reflects the strength of AFM induced by Si. The fact that the difference is large even in fields as high as 20kOe implies that the AFM is very dominant.

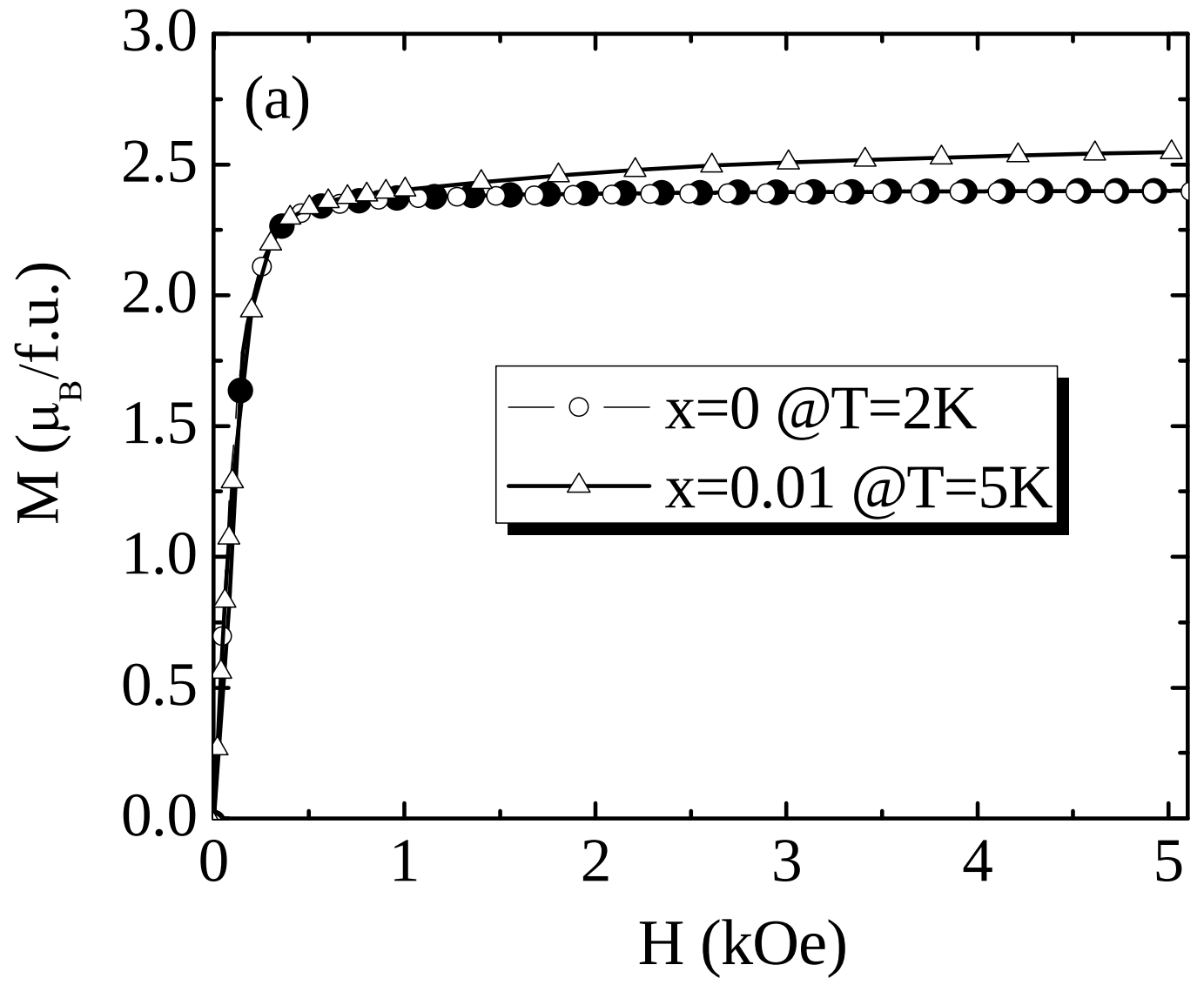

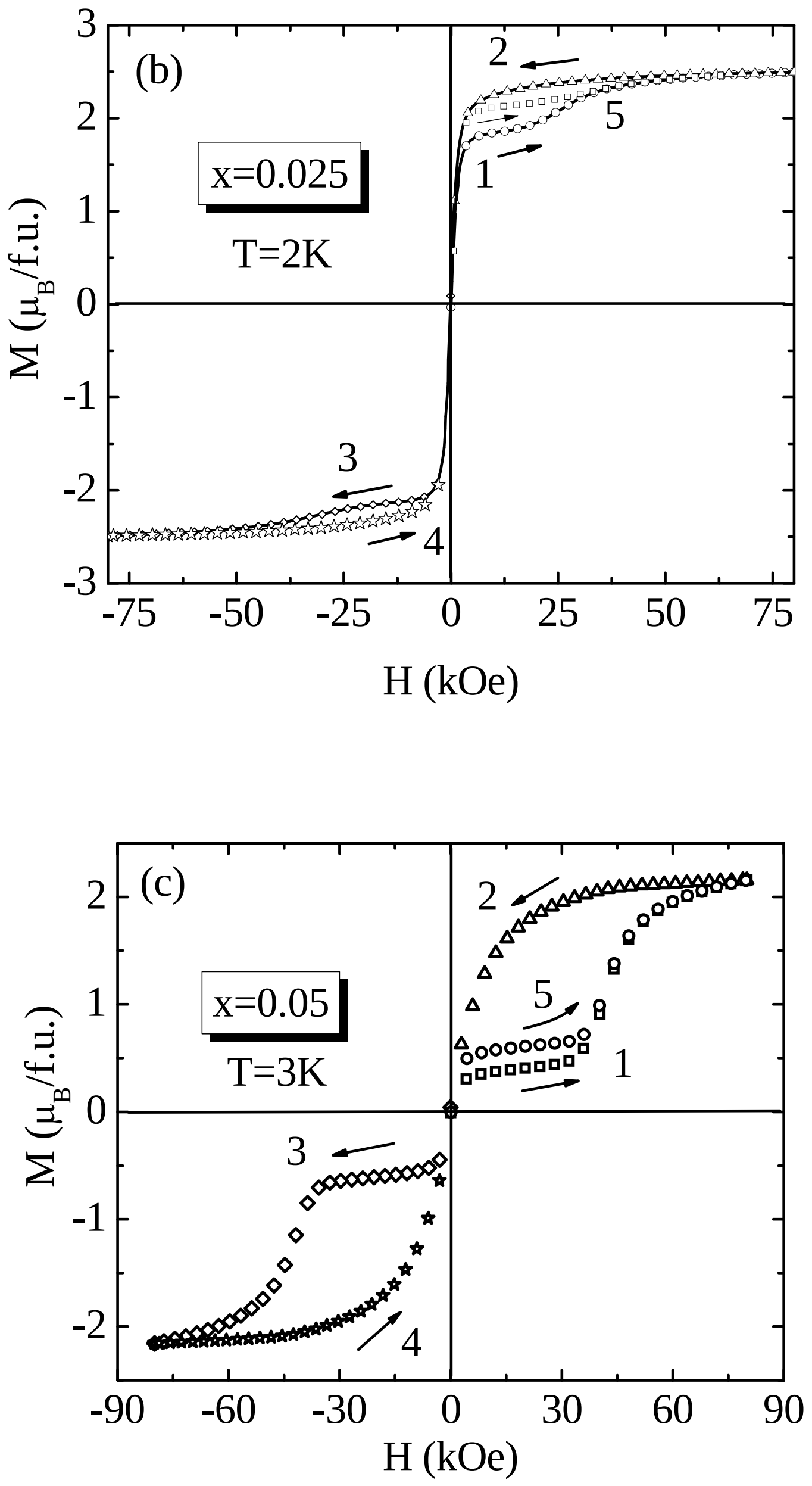

FIG. 3. Isothermal magnetization curves for (a) $Ce(Fe_{1.99}Si_{0.01})_2$ at T=5K, (b) $Ce(Fe_{1.985}Si_{0.025})_2$ at T=2K, (c) $Ce(Fe_{1.95}Si_{0.05})_2$ compound at T=3K. Arrows show the field (Sweep rate: 100Oe/sec) directions in which measurements have been carried out.

Magnetization isotherms have been taken on all the samples at very low temperatures up to maximum field of 90kOe. Fig. 3a shows the M *vs.* H plots for x=0.01 compound along with that of the parent compound. Five loop magnetization data have been taken for

x=0.025 and x=0.05 substituted compounds (Fig. 3b and c). The field sweep rate during all these measurements in fig. 3 was 100 Oe/sec. and the sample was zero field cooled from 325K. M(H) behaviors of x=0 & 0.01 compounds are similar to a typical ferromagnet. Interestingly a metamagnetic transition from AFM to FM phase is observed for higher Si substituted (x=0.025 & 0.05) compounds which was expected after observing the temperature dependence of magnetization data of these compounds. M *vs.* T data also shows that the unstable AFM phase gets partially stabilized in x=0.025 compounds and it almost fully stable with broad AFM ground state near low temperature region for x=0.05 compound. At T $\leq$3K both the compounds (x=0.025 & 0.05) are more or less in AFM dominating ground state. Application of externally applied field favors the final state to be ferromagnetic. Depending on the strength of AFM, the metamagnetic transition occurs with increase in field. The metamagnetic transition in x=0.025 starts at a small field (~ 4kOe) and then the moment increases very rapidly. Then up to a field of about 18kOe moment does not increases much. Further increment in field assists the system to transfer into a fully ferromagnetic phase. Moment does not increase considerably up to 32kOe in x=0.05 compound and a sudden jump in magnetization is observed at ~ 32kOe, aligning the moments parallel to the direction of applied field (Fig. 3c).

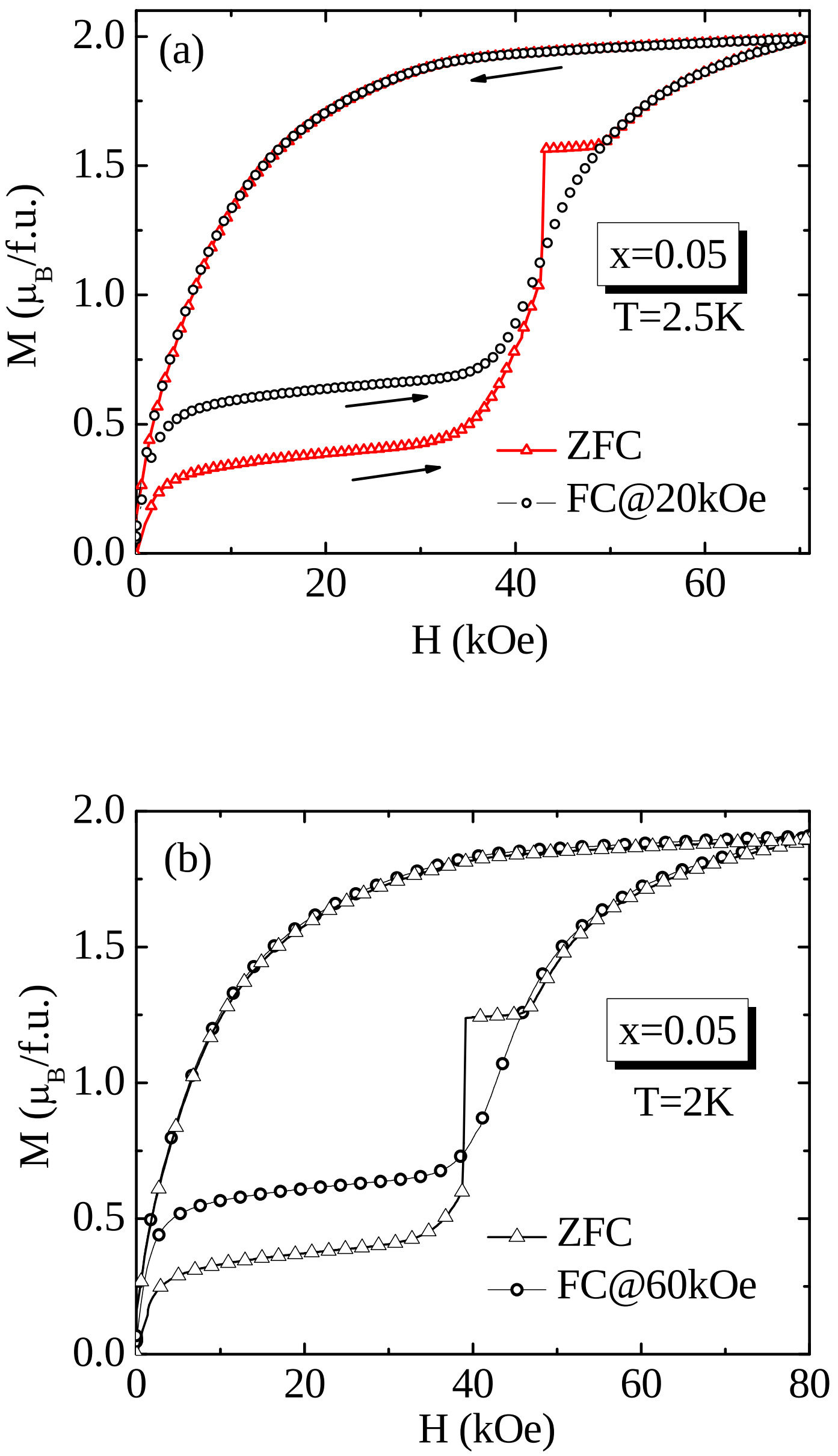

FIG. 4. Isothermal magnetization curve at (a) T=2.5K and (b) T=2K for $Ce(Fe_{1.95}Si_{0.05})_2$ compound. Measurements have been performed in both increasing and decreasing field (sweep rate=100Oe/sec) and the sample was cooled from 325K in zero and 60kOe (a) & 20kOe (b) field.

When we further decrease the temperature from 3K, one can observe dramatic change in magnetization behavior at the AFM - FM transition (Fig. 4). An extremely sharp step is found in M(H) at T=2.5K at a critical field ($H_C$) 42.6kOe while the value of $H_C$ is 38.8kOe at T=2K, for x=0.05n compound. Moment does not increase much beyond this critical field and a plateau region in M(H) is observed which is followed by a sudden jump in moment value. The height of the step ($\Delta \boldsymbol{M}$) is 0.5 and 0.64 $\mu_B / f.u.$ at T=2.5 and 2K respectively. We have found that by lowering the temperature, sharper steps can be realized in this compound and the critical field for appearing the steps is less at T=2K compared to T=2.5K. This critical field is not a unique quantity and depends on the extrinsic measurement protocol which will be shown in the following paragraph. This kind of step behavior was also observed in $Ce(Fe_{1.975}Ga_{0.025})_2$ compound [4] and Ru & Re doped compounds [3]. A similar behavior is also found in other systems like manganites and $Gd_5Ge_4$ [12]. The sharpness of the steps in magnetization in the present case is quite surprising since it is polycrystalline in nature. This is a new sort of metamagnetism which have been often explained by relief of strain across phase separated region [1-4] or burst like growth of FM phase with increasing field [5,6] or quenched disorder [7]. The presence of sharp step in magnetization indicates that this may be a martensitic type behavior across the order-order (AFM–FM) transition.

The data has been also taken when the sample was cooled in a field (both higher and lower than $H_C$). While cooling the sample in 20kOe ($<$ $H_C$=42.6kOe) field to T=2.5K, the low field magnetization got enhanced and a smooth metamagnetic transition occurs across AFM-FM transformation rather than a sharp step as found in ZFC. Even a cooling field 60kOe ($>$ $H_C$=38.8kOe) at T=2K can not convert the system considerably to its FM phase but results a smooth metamagnetic transition. Comparing the situation in the case of Ga doping [4], it is seen that the ground state AFM coupling is stronger in the case of Si doping.

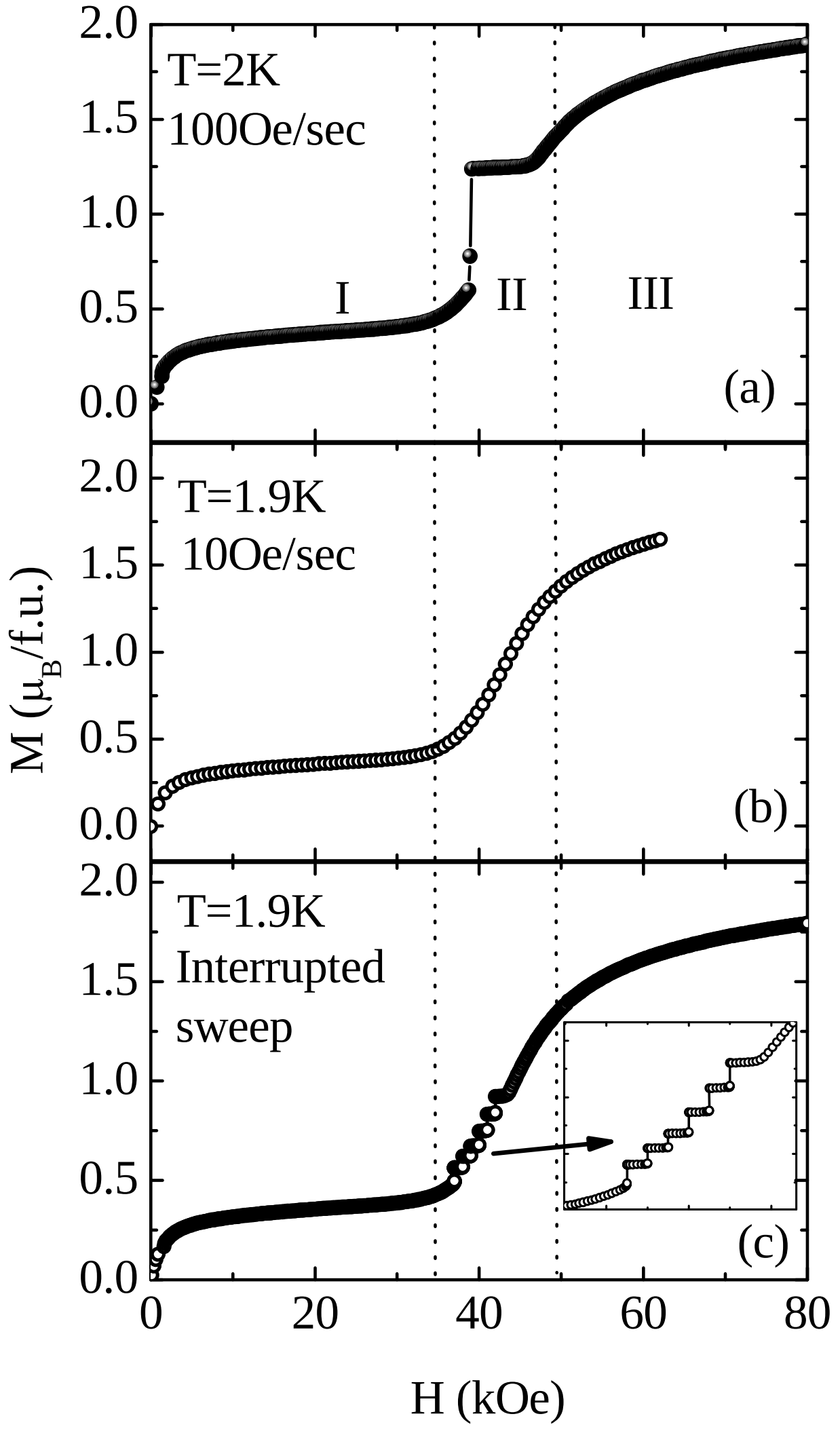

FIG. 5. Isothermal magnetization curves showing the dependence of measurement procedure. Field sweep rates are (a) 100Oe/sec, (b) 10Oe/sec and (c) interrupted sweep.

The formation of FM phase by this discontinuous phase transformation requires the nucleation of the new phase in highly localized regions (special sites) on which the heterogeneous nucleation takes place. The small doping concentration inside the matrix can act as one of these special sites. Nucleation also occurs at the grain boundaries, grain edges and grain corners in a polycrystalline compound. As shown in above figure, stage I is the incubation period in which the AFM phase is metastable. New clusters of very small sizes, which are precursors to the final stable FM phase continuously form and decompose in the matrix. The distribution of these small clusters evolves with time to produce larger clusters which are more stable and therefore less likely to revert back to the matrix. It is observed that when we decrease magnetic field magnetization does not follow the same path. This may be understood by assuming that some of the largest of these clusters evolve into stable nuclei of FM phase and remain in the system permanently and continue to grow with driving force. In stage II, the distribution of the FM clusters has built up into a quasi-steady state. But dramatic observations were found across this transition producing extremely sharp steps to reach the quasi-steady state. In stage III, rate of nucleation is decreased and formation of new stable phase is almost complete.

We have also shown here that the steps can be induced or can be masked by the extrinsic measurement protocol. It can be seen here that definition of critical field totally depends on the extrinsic factors. In above Fig.5, M(H) data has been taken in three different ways; (a) 100 Oe/sec at T=2K, (b) 10 Oe/sec at T=1.9K and (c) interrupted sweep at T=1.9K where the scan was delayed for 2 hrs each at different fields. A sharp step observed in case (a) at 2K, becomes smooth at slower sweep rate even when the temperature is reduced from 2 to 1.9 K. Six small and sharp steps are observed in the interrupted sweep. The absence of steps in slow field sweep rate can be explained from the martensitic picture. A slow change in driving forces (magnetic field) assists the system to transform slowly and progressively from a distorted phase to another more ordered phase. During this process the system finds enough time to overcome the elastic energy smoothly across the two phases which blocks the transition up to a certain field. The influence of the external driving force confirms our presumption that this system posses a martensitic type behavior.

Many doped $CeFe_2$ systems [10, 11] are found to posses a structural distortion across first order AFM-FM transition. We can assume that when clusters form in the matrix phase, an elastic-misfit strain energy is generated because of volume or/and shape incompatibilities between the cluster and the matrix. This strain energy acts as a barrier to the nucleation [13]. The application of magnetic field at low temperature favors the FM phase, in a way, a new crystal structure. Externally applied magnetic field favors FM fraction within the matrix and system releases its strain energy across the transition. It can be assumed here that depending on the experimental condition this release of strain energy may occur in a “military” fashion. As a consequence moments suddenly jump to certain magnetization value producing ultra sharp steps across this order-order transition.

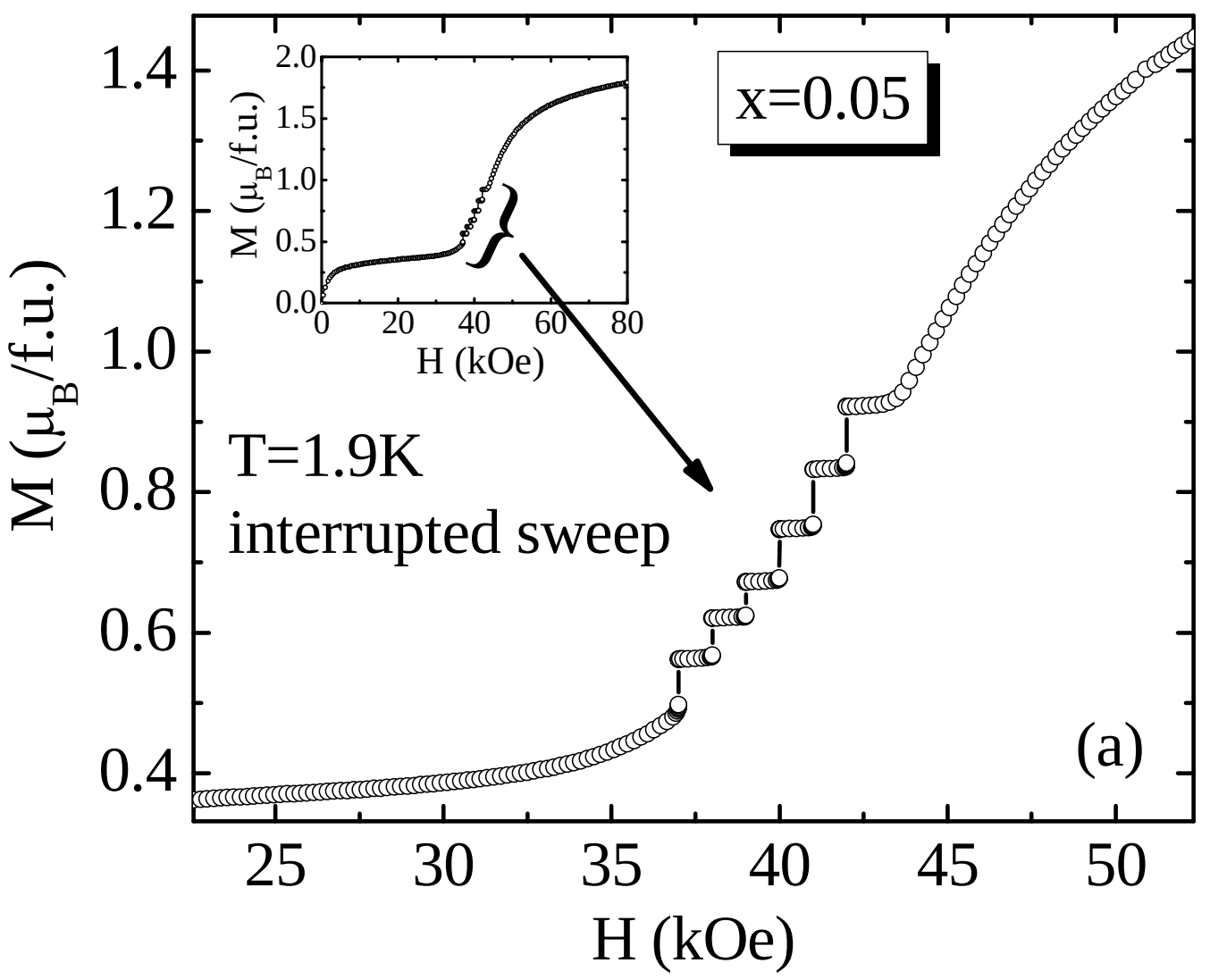

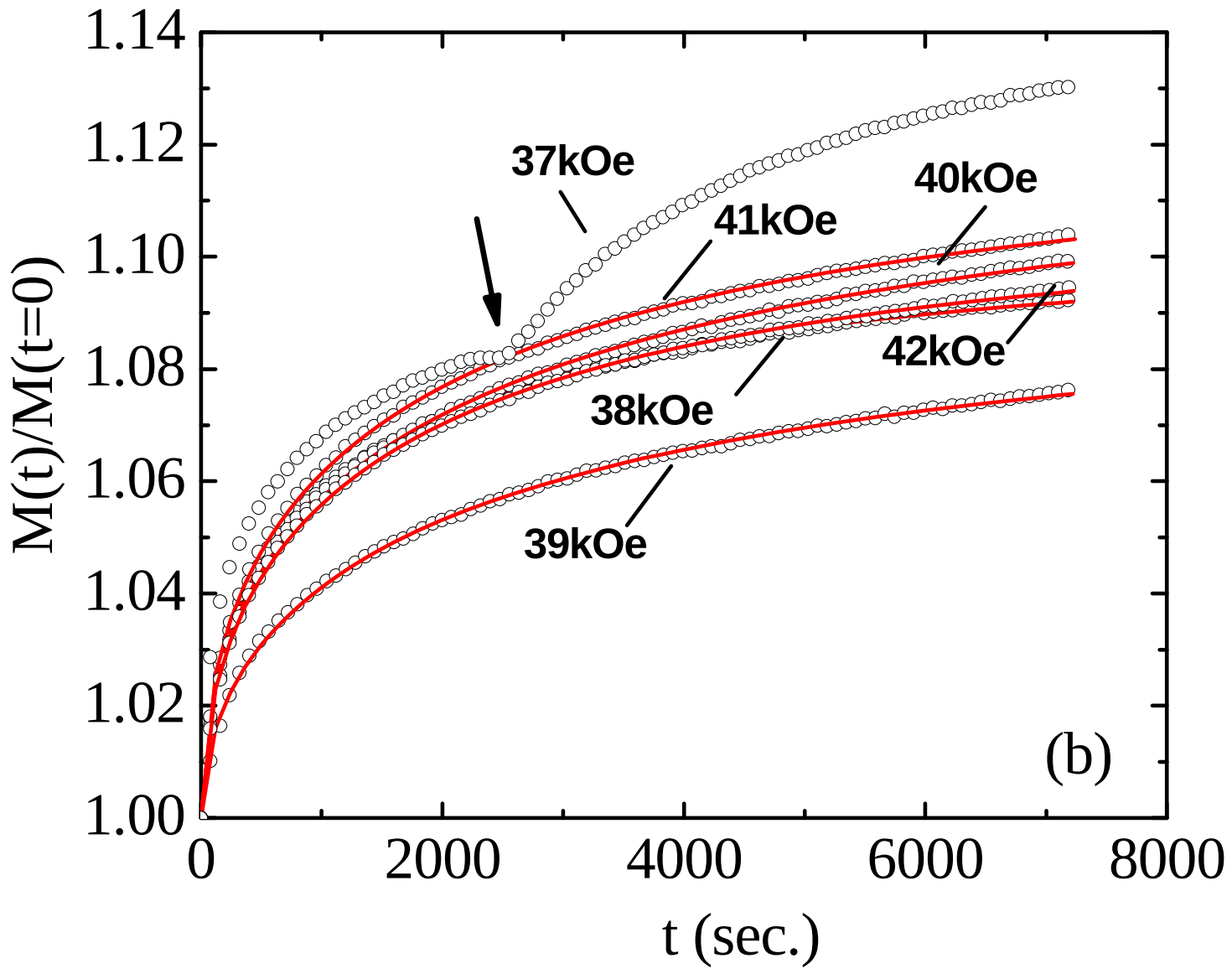

FIG. 6. (a) Field dependence of magnetization at T=1.9K for $Ce(Fe_{1.95}Si_{0.05})_2$ compound. During sweeping from 0 to 80kOe the following fields: 37, 38, 39, 40, 41 and 42 kOe; were held for 2hrs each. Field sweep rate was 100Oe/sec. (b) Variations of normalized magnetization with time at different holding fields. Arrow shows the jump in normalized magnetization for the holding field of 37kOe.

The interrupted sweep and growth of FM phase during those times are demonstrated in Fig. 6. In the interrupted sweep the field sweep rate was 100 Oe/sec. and the following fields (near about the metamagnetic transition region); 37, 38, 39, 40, 41 & 42kOe, were held constant for two hrs. Interestingly six steps are found (Fig. 6a) at each of the above mentioned six fields indicating that there exists an induction period to appear these steps. This type of behavior can be compared with the report by Wu et al. [14]. Magnetic relaxation at different constant fields can be well described by a stretched exponential of this type: $\boldsymbol{M}_{\boldsymbol{t}}(\boldsymbol{H}) = \boldsymbol{M}_0(\boldsymbol{H}) + [\boldsymbol{M}_\infty(\boldsymbol{H}) - \boldsymbol{M}_0(\boldsymbol{H})][1-\exp\{-(\boldsymbol{t}/\tau)^\alpha\}]$. Where $\tau$ is the characteristic relaxation time and $\alpha$ is called stretching parameter that can range between 0 and 1. From our fitting we have obtained the value of the stretching parameter is almost 0.5 and system having a characteristic relaxation $(\tau)$ in a range of 1300 to 2700 sec. depending on the applied field. System relaxes to its actual magnetization value for a particular field during these holding times. The growth of the FM phase at different fields has been shown in Fig. 6b. A jump in normalized magnetization is observed for a holding field of 37kOe (arrow in Fig. 6b). The moments are relaxed to different final values for different holding fields. As a consequence the step sizes in M(H) depend on the holding fields and also on the wait time (here 2 hrs. for each holding field).

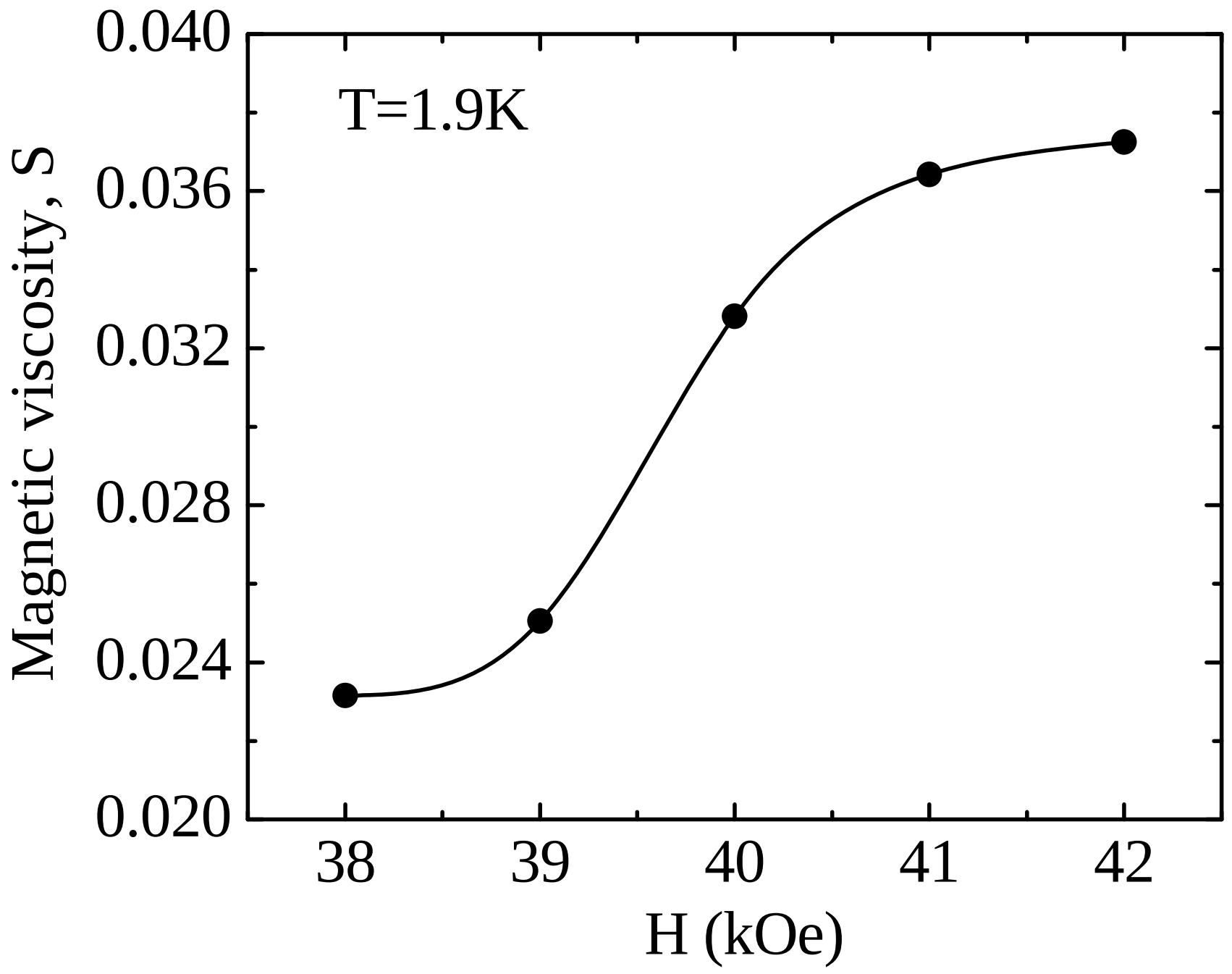

FIG. 7. Field dependence of magnetic viscosity for x=0.05 compound at T=1.9K.

Magnetic viscosity has been calculated using the relation: $\boldsymbol{S} = (1/\boldsymbol{M}_0) \times \{\boldsymbol{dM}/\boldsymbol{d}(\log_{10} \boldsymbol{t})\}]$. At 2K the metamagnetic type transition starts around 36 kOe and up to nearly 52 kOe there is a huge change in magnetization is observed (see Fig. 4b). The change of magnetic viscosity (S) across this region has been shown in Fig. 7. An increment S with increasing the field can be clearly observed from this figure. At higher field (above 41kOe), S seems to have almost saturated value. While there is a huge jump of S is found around the field 39-41 kOe region. We conclude here that the observed variation of S is consistent with the field and can attributed to the change in fraction of antiferro and ferromagnetic phases.

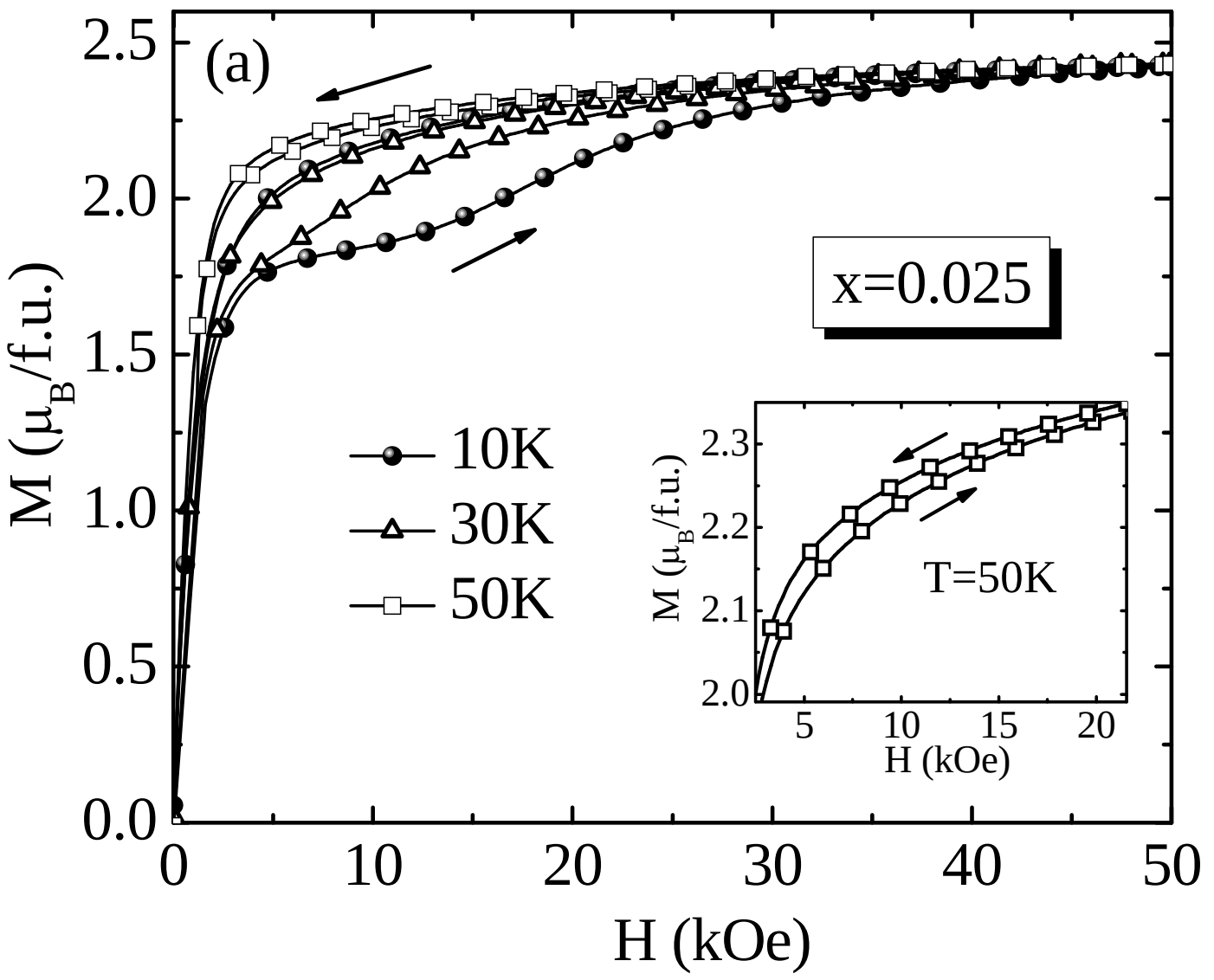

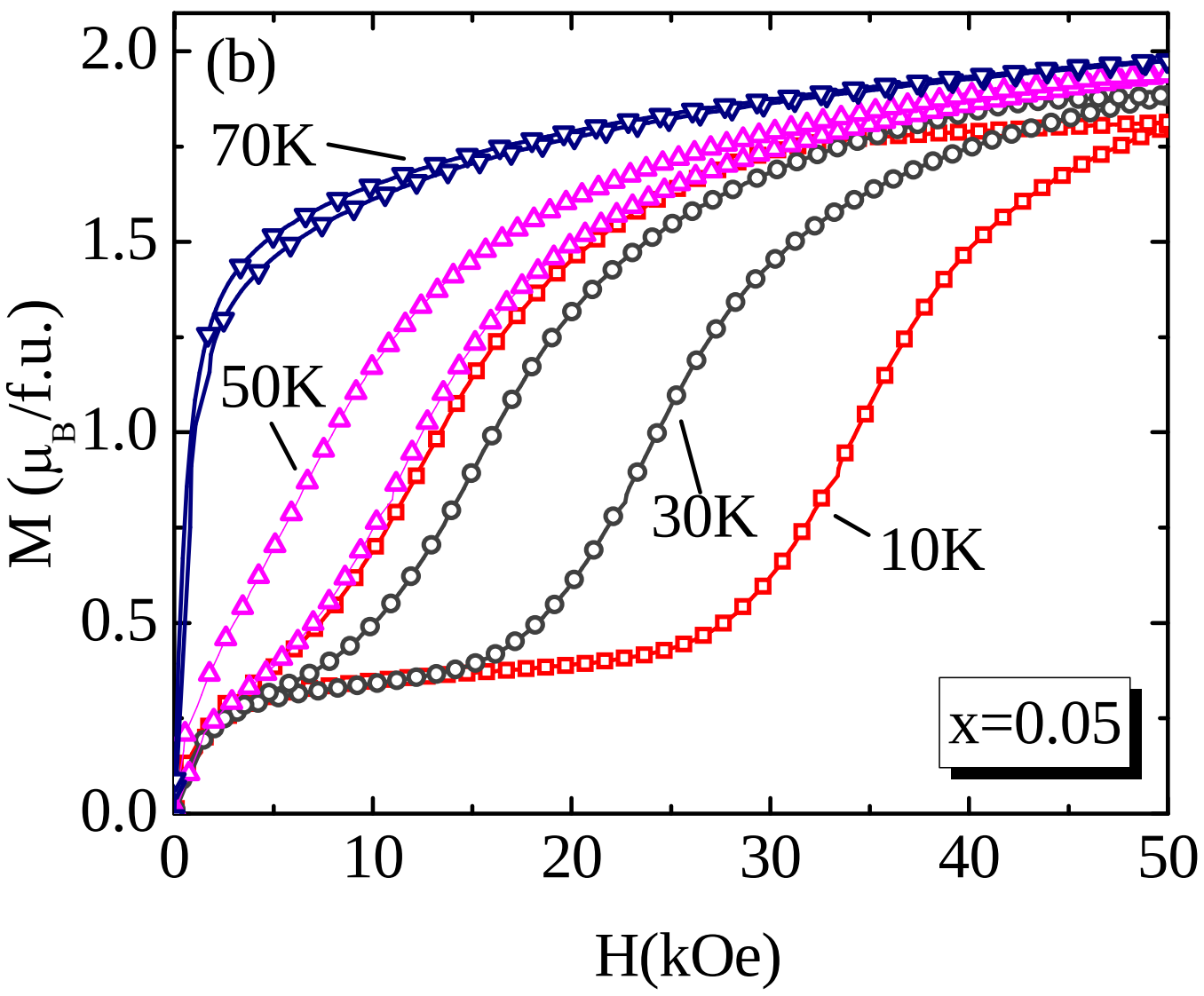

FIG. 8. Isothermal magnetization at various higher temperatures for (a) $Ce(Fe_{1.985}Si_{0.025})_2$ and (b) $Ce(Fe_{1.95}Si_{0.05})_2$ compounds. Area of the hysteresis loops getting diminished with increment in temperature.

Fig. 8 shows the high temperature behavior of magnetization isotherms of both x=0.025 and 0.05 compounds. Arrows indicate increasing and decreasing field. In both the cases with increase of temperature area of the hysteresis loops gets decreased. Area of the loop can be made to zero above 50K in x=0.025 compound and 70K for x=0.05 compound.

## IV. Conclusions

Detailed magnetization measurements have been performed in Si doped $CeFe_2$ polycrystalline compounds. This work presents the appearance of the steps in the magnetization isotherms, which can be induced by extrinsic experimental protocol. Presence of incubation time and variation of magnetic viscosity has been discussed to understand the dynamics of the system. Sharp steps across field induced AFM to FM phase have been attributed to the martensitic behavior during transformation. Multiple steps can be achieved with proper relaxation protocol. Reproducibility of the steps and strong dependence on external parameters open up lots of opportunity to study the dynamics in these kinds of systems. This study also shows that there are several similarities between Ga and Si doping in $CeFe_2$ with regard to the martensitic scenario and magnetization behavior.

## V. References

[1]R. Mahendiran, A. Maignan, S. Hebert, C. Martin, M. Hervieu, B. Raveau, J.F. Mitchell, and P. Schiffer, Phys. Rev. Lett. **89**, 286602 (2002).

[2]V. Hardy, S. Majumdar, S. J. Crowe, M. R. Lees, D. McK. Paul, L. Herve, A. Maignan, S. Hebert, C. Martin, C. Yaicle, M. Hervieu, and B. Raveau, Phys. Rev. B **69**, 020407(R) (2004).

[3]S. B. Roy, M. K. Chattopadhyay, and P. Chaddah, Phys. Rev. B **71**, 174413 (2005).

[4] Arabinda Haldar, K. G. Suresh and A. K. Nigam, Phys. Rev. B (In Press).

[5]V. Hardy, A. Maignan, S. Hebert, C. Yaicle, C. Martin, M. Hervieu, M. R. Lees, G. Rowlands, D. M. Paul, and B. Raveau, Phys. Rev. B **68**, 220402(R) (2003).

[6]V. Hardy, S. Hébert, A. Maignan, C. Martin, M. Hervieu, and B. Raveau, J. Magn. Magn. Mater. **264**, 183 (2003).

[7]L. M. Fisher, A. V. Kalinov, I. F. Voloshin, N. A. Babushkina, D. I. Khomskii, Y. Zhang and T. T. M. Palstra, Phys. Rev. B, **70**, 212411 (2004).

[8]Olle Eriksson, Lars Nordstrom, M.S.S. Brooks, Borje Johansson, Phys. Rev. Lett., **60**, 2523 (1988).

[9]C. Giorgetti, S. Pizzini, E. Dartyge, A. Fontaine, F. Baudelet, C. Brouder, P. Bauer, G. Krill, S. Miraglia, D. Fruchart and J. P Kappler, Phys. Rev. B **48,** 12732(1993).

[10]S. B. Roy and B. R. Coles, J. Phys.: Condens. Matter **1**, 419 (1989).

[11]S. B. Roy and B. R. Coles, Phys. Rev. B **39**, 9360 (1990).

[12]E.M. Levin, K.A. Gschneidner, Jr., and V.K. Pecharsky, Phys. Rev. B **65**, 214427 (2002).

[13]Robert W. Balluffi, Samuel M. Allen and W. Craig Carter, *Kinetics of Materials* (John Wiely & Sons, Inc.)

[14]T. Wu and J. F. Mitchell, Phys. Rev. B, **69**, 100405(R) (2004).